\documentclass[11pt]{amsart}
\usepackage[letterpaper,margin=1.08in]{geometry}
\usepackage{amssymb}
\usepackage[svgnames]{xcolor}
\usepackage{hyperref}
\hypersetup{colorlinks=true,allcolors=Navy,breaklinks=true,
  pdftitle={There is no binary linear [65,9,29] code},
  pdfauthor={Zhichao Tang, Xiang Fan}}
\newtheorem{theorem}{Theorem}[section]
\newtheorem{lemma}[theorem]{Lemma}
\newtheorem{proposition}[theorem]{Proposition}
\numberwithin{equation}{section}
\title{There is no binary linear $[65,9,29]$ code}
\author{Zhichao Tang}
\author{Xiang Fan}
\address{School of Mathematics, Sun Yat-sen University}
\date{}
\begin{document}
\begin{abstract}
We prove that no binary linear $[65,9,\geqslant29]$ code exists, equivalently, that no even binary linear $[66,9,\geqslant30]$ code exists. Together with known constructions, this determines $d_2(65,9)=28$, $n_2(9,29)=66$, and $n_2(9,30)=67$. A residual-code bound and a known construction also give $d_2(122,10)=56$. To prove nonexistence, we associate with a hypothetical even code a quadratic function derived from half the codeword weight modulo $2$. Its polar form determines the fourth derivative of a Boolean function of degree at most four. Character sums, weight moments, and the exclusion of weight-$42$ codewords give upper bounds on the support of this function in seven cases. Restrictions to nondegenerate affine subspaces exclude six cases. In the remaining case, equality would force four distinct coordinate functionals to be identically $1$ on an affine seven-space, although at most two can have this property. The argument uses published minimum-length results for binary linear codes of dimensions eight and nine.
\end{abstract}
\maketitle

\section{Introduction}

A \emph{binary linear code} of length $n$ and dimension $k$ is a $k$-dimensional subspace of $\mathbb{F}_2^n$. The \emph{Hamming weight} of a word is its number of nonzero coordinates. The \emph{minimum distance} of a code is the smallest weight of a nonzero codeword. We write $[n,k,d]$ for a code with minimum distance exactly $d$ and $[n,k,\geqslant d]$ for one with minimum distance at least $d$. A code is \emph{even} if all its words have even weight. Write $n_2(k,d)$ for the smallest length of a binary linear code of dimension $k$ and minimum distance $d$, and $d_2(n,k)$ for the largest possible minimum distance of a binary linear code of length $n$ and dimension $k$.

Determining these extremal parameters is a basic problem in coding theory. Building on minimum-length results for dimensions eight and nine~\cite{BJV00,BGV00}, we resolve the existence question for a binary $[65,9,29]$ code identified as the next open case by Kurz~\cite[discussion following Theorem~3.3]{Kurz21}. Recent work gives bounds for nine-dimensional \emph{self-orthogonal} binary codes, in which every pair of codewords has zero inner product~\cite{Kurz26}; the theorem below imposes no such condition.

\begin{theorem}\label{thm:main}
There is no binary linear $[65,9,\geqslant29]$ code. Equivalently, there is no even binary linear $[66,9,\geqslant30]$ code.
\end{theorem}

To see the equivalence, \emph{puncturing} means deleting the same coordinate from every codeword. Adjoining the sum of the $65$ coordinates as a \emph{parity coordinate} to a binary $[65,9,\geqslant29]$ code gives an even $[66,9,\geqslant30]$ code. Conversely, puncturing such an even code in any coordinate produces a binary $[65,9,\geqslant29]$ code: its dimension remains $9$, and its minimum distance decreases by at most one.

Cossidente and Siciliano constructed a binary linear $[67,9,30]$ code~\cite[Section~IV]{CS01}. Puncturing it once gives a $[66,9,\geqslant29]$ code, and puncturing it twice gives a $[65,9,\geqslant28]$ code; both have dimension $9$. Theorem~\ref{thm:main} forces the twice-punctured code to have distance $28$. The once-punctured code has distance $29$: otherwise puncturing it again would yield a forbidden $[65,9,\geqslant29]$ code. Consequently,
\[
 d_2(65,9)=28,\qquad n_2(9,29)=66,\qquad n_2(9,30)=67.
\]
Combining a residual-code bound with a construction recorded by Grassl~\cite{GrasslTables} also determines $d_2(122,10)=56$ (Proposition~\ref{prop:dimension10}).

\subsection*{Proof outline}
Suppose an even $[66,9,\geqslant30]$ code exists, with message space
$V:=\mathbb{F}_2^9$ and codeword weights $w(x)$. A \emph{Boolean function}
on $V$ is a map from $V$ to $\mathbb{F}_2$; its \emph{Boolean degree} is
the degree of its unique squarefree polynomial representation. Write
$D_uF(x):=F(x+u)+F(x)$ for its \emph{derivative} in direction $u$.
Let $q(x)$ be $w(x)/2$ modulo $2$. This is a \emph{quadratic function},
of Boolean degree at most two. Its \emph{polar form},
$(x,y)\mapsto q(x+y)+q(x)+q(y)$, is the inner product of the
corresponding codewords. Let $h(x)$ be $(w(x)+2q(x))/4$ modulo $2$,
with $q(x)$ represented by $0$ or $1$ in the numerator.
This function has degree at most four, and its fourth derivative is
determined by the same polar form.

The \emph{character sums} $\sum_{x\in V}(-1)^{q(x)+\ell(x)}$,
where $\ell$ is a linear functional, and the first three \emph{weight moments}
$\sum_{x\in V}w(x)^j$, $j=1,2,3$, together with the exclusion of
weight-$42$ words, leave seven parameter configurations and give upper bounds
on the size of $\{x:h(x)=1\}$ (Lemma~\ref{lem:types} and
Proposition~\ref{prop:population}). The fourth-derivative identity
forces $h$ to have odd weight on nondegenerate affine four-spaces;
affine slicing and Parseval's identity then exclude six cases
(Lemma~\ref{lem:lowerbounds}). In the last case, equality forces four
distinct coordinate functionals to be identically $1$ on an affine
seven-space. Its direction space has codimension two, so at most two
linear functionals can have this property.

The use of Boolean functions and Reed--Muller weight restrictions in
coding theory has a precedent in Simonis~\cite{Simonis94}; the present
argument uses the specific fourth-polarization identity and its equality
conditions. It also relies on three published minimum-length bounds,
stated in Section~\ref{sec:reductions}. The first two are
computer-assisted~\cite{BJV00,BGV00}; the third is a dimension-eight
table result supported by the exclusion of a binary $[25,8,10]$ code~\cite{YH90}.

\section{Cited bounds and the quadratic function}\label{sec:reductions}

We use the following published bounds:
\begin{equation}\label{eq:external}
\begin{gathered}
\text{no binary }[64,8,\geqslant30]\text{ code},\\
\text{no binary }[n,9,\geqslant32]\text{ code with }n\leqslant69,\\
\text{no binary }[n,8,\geqslant10]\text{ code with }n\leqslant25.
\end{gathered}
\end{equation}
The first is Bouyukliev--Jaffe--Vavrek~\cite[Theorem~4]{BJV00}. The second follows from $n_2(9,32)=70$ in Bouyukliev--Guritman--Vavrek~\cite[Theorem~2]{BGV00}. The \emph{Griesmer bound} for binary linear codes reads $n_2(k,d)\geqslant g_2(k,d)$, where
$g_2(k,d):=\sum_{i=0}^{k-1}\left\lceil d/2^i\right\rceil$.
The table on page~1539 of~\cite{BJV00} gives $n_2(8,10)-g_2(8,10)=2$. Since
\[
 g_2(8,10)=10+5+3+2+1+1+1+1=24,
\]
it follows that $n_2(8,10)=26$, proving the third bound. The nonexistence of a binary $[25,8,10]$ code was established by Ytrehus and Helleseth~\cite{YH90}. None of these bounds requires the code to be even. The cited results concern exact minimum distance. To deduce the formulations in~\eqref{eq:external}, take a code with minimum distance $\delta\geqslant d$ and delete $\delta-d$ positions in a word of minimum weight. The resulting code still has dimension $k$ and has minimum distance exactly $d$. Zero coordinates may then be appended to reach a prescribed larger length.

Assume henceforth that $C$ is an even binary $[66,9,\geqslant30]$ code. Choose a linear isomorphism $c\colon V\longrightarrow C$, where $V=\mathbb{F}_2^9$, and let $\ell_1,\ldots,\ell_{66}\in V^*$ be its coordinate functions. Set $w(x):=\operatorname{wt}(c(x))$.

\begin{lemma}\label{lem:columns42}
The coordinate functions $\ell_i$ are distinct and nonzero. Moreover, $C$ has no word of weight $42$.
\end{lemma}
\begin{proof}
If one coordinate function is zero, restrict $V$ to the kernel of a nonzero coordinate function and delete these two zero coordinates. This gives a binary $[64,8,\geqslant30]$ code. If two nonzero coordinate functions coincide, restrict to their common kernel and delete both coordinates, with the same result. Both possibilities contradict~\eqref{eq:external}.

Suppose that $w(u)=42$. Project $C$ onto the $24$ coordinates outside the support of $c(u)$. The kernel is exactly $\langle c(u)\rangle$: otherwise two nonzero words $c(x)$ and $c(x+u)$ would both be supported on these $42$ positions and would have total weight $42<60$. Thus the projected code has dimension $8$. If its word corresponding to $x\notin\langle u\rangle$ has weight $b$, then
\[
 w(x)+w(x+u)=42+2b\geqslant60,
\]
so $b\geqslant9$. Adjoining its parity coordinate gives a binary $[25,8,\geqslant10]$ code, again contrary to~\eqref{eq:external}.
\end{proof}

Define $q\colon V\to\mathbb{F}_2$ and its polar form $B$ by
\[
 q(x):=\frac{w(x)}2\pmod2,
 \qquad B(x,y):=q(x+y)+q(x)+q(y)=c(x)\mathbin{\cdot}c(y).
\]
Thus $B$ is alternating and bilinear. Let $R:=\{x\in V:B(x,y)=0\text{ for every }y\in V\}$ be its \emph{radical}, and put
\[
 r:=\dim R,\qquad s(x):=(-1)^{q(x)},\qquad
 T:=\lvert\{x:q(x)=0\}\rvert,\qquad S:=2T-512.
\]
The rank of an alternating form is even, so $r\in\{1,3,5,7,9\}$. The restriction of $q$ to $R$ is linear. Whenever $q(x)$ occurs in an integer expression, it stands for its representative in $\{0,1\}$.

For $\ell\in V^*$, set
\[
 G(\ell):=\sum_{x\in V}s(x)(-1)^{\ell(x)},\qquad
 \mathcal{A}:=\{\ell\in V^*: \ell|_R=q|_R\},\qquad
 a:=2^{(7+r)/2},\qquad b:=2^{(9-r)/2}.
\]
Character orthogonality gives
\begin{equation}\label{eq:gauss}
 G(\ell)^2=512\sum_{z\in R}(-1)^{q(z)+\ell(z)}.
\end{equation}
Indeed, in the square of $G(\ell)$, substitute $x+z$ for the second variable and sum first over $x$. Hence $G$ vanishes outside $\mathcal{A}$ and equals $\pm2a$ on $\mathcal{A}$. In particular, if $q|_R\ne0$, then $S=0$; otherwise $S=\pm2a$. When $r=9$ and $q|_R=0$, only $S=512$ occurs.

Write
\[
 \tau_\ell:=\frac{G(\ell)}{2a}\quad(\ell\in\mathcal{A}),\qquad
 Z:=\{\ell_1,\ldots,\ell_{66}\}\cap\mathcal{A},\qquad
 m:=\lvert Z\rvert,\qquad \varepsilon:=\sum_{\ell\in Z}\tau_\ell.
\]
By Lemma~\ref{lem:columns42}, $Z$ is a set, not a multiset. Fourier inversion and the character expression for the weight give
\begin{equation}\label{eq:gaussweight}
 \sum_{\ell\in\mathcal{A}}\tau_\ell(-1)^{\ell(x)}=b\,s(x),
 \qquad
 \Xi:=\sum_{x\in V}s(x)w(x)=33S-a\varepsilon.
\end{equation}
For a subspace $U\leq V$, let $e(U)$ denote the number of coordinates whose restrictions to $U$ are nonzero. Each such coordinate takes the value $1$ on half of $U$, so
\begin{equation}\label{eq:supportaverage}
 \sum_{u\in U}w(u)=2^{\dim U-1}e(U),\qquad
 \frac1{\lvert U\rvert}\sum_{u\in U}w(x+u)\leqslant66-\frac{e(U)}2.
\end{equation}
The second inequality holds because the remaining $66-e(U)$ coordinates are constant on $x+U$.

\section{Moment identities and upper bounds}\label{sec:weights}

The \emph{weight distribution} of $C$ is given by $A_i:=\lvert\{x\in V:w(x)=i\}\rvert$, and the polynomial $\sum_i A_i z^i$ is its \emph{weight enumerator}. The \emph{dual code} is
$C^\perp:=\{v\in\mathbb{F}_2^{66}:v\mathbin{\cdot}c=0\text{ for all }c\in C\}$;
write $B_i$ for its number of words of weight $i$, and $\beta:=B_3$. Since the coordinate functions are distinct and nonzero, $B_1=B_2=0$ and
\begin{equation}\label{eq:moments}
 \sum_xw(x)=16896,\qquad \sum_xw(x)^2=566016.
\end{equation}
Indeed, a coordinate is $1$ on $256$ messages, and two distinct coordinates are simultaneously $1$ on $128$ messages.

Define the cubic polynomials
\[
 P_0(w):=4(w-32)(w-38)(w-40),\qquad
 P_1(w):=4(w-30)(w-38)(w-42).
\]
For $x\ne0$, the value $P_{q(x)}(w(x))$ is nonnegative, since $w(x)\geqslant30$ and $w(x)\equiv2q(x)\pmod4$. For length $66$, the binary \emph{Krawtchouk polynomials} of degrees $1$, $2$, and $3$ are
\[
 \begin{aligned}
 K_1(w)&=66-2w,\\
 K_2(w)&=2145-132w+2w^2,\\
 K_3(w)&=45760-\frac{12872}{3}w+132w^2-\frac43w^3.
 \end{aligned}
\]
Here $K_j(w(x))$ is the sum of $(-1)^{\sum_{i\in I}\ell_i(x)}$ over all $j$-subsets $I$ of the coordinate positions. Consequently $\sum_xK_j(w(x))=512B_j$. Expanding the cubics gives
\[
 P_{q(x)}(w(x))=1134-3040\,\mathbf{1}_{q(x)=0}
 +40s(x)w(x)-124K_1(w(x))-22K_2(w(x))-3K_3(w(x)).
\]
Since $P_0(0)=-194560$, summation yields
\begin{equation}\label{eq:cubic}
 \sum_{x\ne0}P_{q(x)}(w(x))
 =40\Xi-3040T+775168-1536\beta.
\end{equation}
In particular, the right side is nonnegative.

\begin{lemma}\label{lem:types}
The values of $(r,q|_R,S)$ and the corresponding upper bounds on $m$ are among the following seven possibilities:
\begin{equation}\label{eq:types}
\begin{array}{c|c|r|r}
 r&q|_R&S&m\ \text{at most}\\ \hline
 5&0&-128&4\\
 3&0&-64&10\\
 1&0&-32&34\\
 1&0&32&34\\
 1&\ne0&0&50\\
 3&\ne0&0&18\\
 5&\ne0&0&6
\end{array}
\end{equation}
\end{lemma}
\begin{proof}
First, no subspace $H\leq V$ of dimension $h\geqslant7$ can have $q|_H=0$. To see this, append to the code the coordinates of a quotient map $V\to\mathbb{F}_2^{9-h}$ and their sum. A nonzero word in $H$ already has weight at least $32$. A word outside $H$ gains a nonzero even vector, of weight at least $2$. This produces a binary code of dimension $9$, length at most $69$, and minimum distance at least $32$, contradicting~\eqref{eq:external}.

Thus $r\geqslant7$ is excluded when $q|_R=0$. If $q|_R\ne0$, put $R_0=\ker(q|_R)$ and choose $r_1\in R$ with $q(r_1)=1$. A complement $Y$ of $R$ has nondegenerate polar form. Choose a subspace $U\leq Y$ of dimension $(9-r)/2$ on which $B$ vanishes. The subspace
\[
 R_0+\{u+q(u)r_1:u\in U\}
\]
has dimension $(r+7)/2$, and $q$ vanishes on it. Hence the possibilities $r=7,9$ are excluded in this case as well.

When $q|_R=0$, every nonzero radical word has weight at least $32$, so~\eqref{eq:supportaverage} gives
\begin{equation}\label{eq:radicalm}
 e(R)\geqslant64-2^{6-r},\qquad
 m=66-e(R)\leqslant2+2^{6-r}.
\end{equation}
Suppose now that $S>0$ and $r\geqslant3$. Since $2a=S$, equation~\eqref{eq:gaussweight} implies
\[
 \Xi\leqslant S(33+m/2)\leqslant S(34+2^{5-r})\leqslant38S.
\]
But~\eqref{eq:cubic} gives $40\Xi\geqslant1520S+3072$, a contradiction. Together with~\eqref{eq:gauss}, this leaves the seven cases in~\eqref{eq:types}. The first four bounds on $m$ follow from~\eqref{eq:radicalm}.

If $q|_R\ne0$, then a coordinate that vanishes on $R_0$ restricts to $R$ either as zero or as $q|_R$. Therefore $e(R)=e(R_0)+m\leqslant66$. For $r=3$ and $5$, equation~\eqref{eq:supportaverage} gives $e(R_0)\geqslant48$ and $60$, respectively, proving $m\leqslant18$ and $6$.

It remains to bound $m$ when $S=0$ and $r=1$. Let
$R=\langle p\rangle$, so $q(p)=1$, $m=w(p)$, and
$s(x+p)=-s(x)$. For $x\ne0$ put
$f(x)=(w(x)-32+2q(x))/4$, and let $f(0)=0$.
These are nonnegative integers. Write $g=(w-31-s)/4$, so
$f=g+8\mathbf{1}_{\{0\}}$. For $i\ne j$, the product $\ell_i(x)\ell_j(x+p)$ equals $1$
on exactly $128$ messages. For $i=j$, it equals $1$ on $256$
messages if $\ell_i(p)=0$, and on none otherwise. Since
$w(p)=m$, it follows that
\[
 \sum_xw(x)w(x+p)=128(66\cdot67-2m),\qquad
 \sum_xs(x)s(x+p)=-512,
\]
where the second equality uses $s(x+p)=-s(x)$.
Expanding $g(x)g(x+p)$ and using~\eqref{eq:moments}
and~\eqref{eq:gaussweight}, we obtain
\[
 16\sum_xg(x)g(x+p)=9984-256m+2\Xi.
\]
Passing from $g$ to $f$ adds $4(m-30)$ to the sum
of products, since $g(p)=(m-30)/4$. As $\Xi=-16\varepsilon$,
\[
 0\leqslant\sum_xf(x)f(x+p)=504-12m-2\varepsilon
 \leqslant504-10m.
\]
Thus $m\leqslant50$.
\end{proof}

Define
\[
 h(x):=\frac{w(x)+2q(x)}4\pmod2,\qquad
 E:=\{x\in V:h(x)=1\},\qquad W:=\lvert E\rvert.
\]
The quotient is integral because
$w(x)\equiv2q(x)\pmod4$. Lemma~\ref{lem:columns42} gives $A_{42}=0$.
To state the resulting identity, let
\[
 \begin{aligned}
 \mathcal{E}:={}&96A_{52}+128A_{54}+224A_{56}+276A_{58}\\
 &+416A_{60}+512A_{62}+704A_{64}+836A_{66}.
 \end{aligned}
\]

\begin{proposition}\label{prop:population}
The weight distribution satisfies
\begin{equation}\label{eq:population}
 4W+4A_{34}+16\beta+32A_{48}+36A_{50}+\mathcal{E}
 =-288-\frac{11S}{4}-\frac{3a\varepsilon}{4}.
\end{equation}
Consequently, the corresponding upper bounds for $W$ are
\begin{equation}\label{eq:Wbounds}
\begin{array}{c|c|r|r}
 r&q|_R&S&W\ \text{at most}\\ \hline
 5&0&-128&64\\
 3&0&-64&32\\
 1&0&-32&52\\
 1&0&32&8\\
 1&\ne0&0&78\\
 3&\ne0&0&36\\
 5&\ne0&0&0
\end{array}
\end{equation}
\end{proposition}
\begin{proof}
For $x\ne0$, put $t=w(x)$ and define
\[
 d(x)=4h(x)+4\mathbf{1}_{t=34}+32\mathbf{1}_{t=48}
       +36\mathbf{1}_{t=50}.
\]
The permissible weights satisfy $30\leqslant t\leqslant66$ and
$t\equiv2q(x)\pmod4$; the weight $42$ is excluded by
Lemma~\ref{lem:columns42}. For the permitted weights, let $\Delta(t)=0$ if $t\leqslant50$,
and otherwise let $\Delta(t)$ be the coefficient of $A_t$ in
$\mathcal{E}$. Direct substitution gives
\[
\begin{aligned}
96\bigl(d(x)+\Delta(t)\bigr)
 &=P_{q(x)}(t)-16(t-32+2q(x))(t-40+2q(x))\\
 &=4(t-32+2q(x))(t-40+2q(x))(t-42-4q(x)).
\end{aligned}
\]
The excluded weight $42$ would require $\Delta(42)=-12$,
so Lemma~\ref{lem:columns42} is essential here.
The quadratic factor in the first line is
$t^2-(70+2s(x))t+1210+70s(x)$; by~\eqref{eq:moments}
and~\eqref{eq:gaussweight}, its sum over $x\ne0$ is
$1536-2\Xi+70S$. Summing the identity, using~\eqref{eq:cubic}
and $\Xi=33S-a\varepsilon$, gives~\eqref{eq:population}.
Since $A_{34},\beta,A_{48},A_{50},\mathcal{E}\geqslant0$,
\[
 W\leqslant-72-\frac{11S}{16}-\frac{3a\varepsilon}{16}.
\]
The bound $\varepsilon\geqslant-m$ and~\eqref{eq:types}
now give~\eqref{eq:Wbounds}.
\end{proof}

\section{Lower bounds from affine restrictions}\label{sec:boolean}

We next obtain lower bounds for the support of $h$ by restricting it to affine subspaces.

Recall that $D_uF(x)=F(x+u)+F(x)$. If $F$ has Boolean degree at most four, its fourth derivative is independent of $x$ and defines the \emph{fourth polarization}.

\begin{lemma}\label{lem:polarization}
The function $h$ has Boolean degree at most four, and
\begin{equation}\label{eq:polarization}
 D_uD_vD_zD_th
 =B(u,v)B(z,t)+B(u,z)B(v,t)+B(u,t)B(v,z).
\end{equation}
It takes the value $1$ an odd number of times on every affine four-dimensional subspace on which $B$ is nondegenerate. If $r_0\in R$, then $D_{r_0}h$ has degree at most two. The same statements hold after restriction to any affine subspace, with its restricted form and its own radical.
\end{lemma}
\begin{proof}
Let $\sigma_j$ denote the elementary symmetric polynomial of degree $j$ in the $66$ code coordinates. The binary digits of an even weight give $q=\sigma_2$ and
\[
 h=\sigma_4+\sigma_2.
\]
To compute the fourth polarization, use the basis $e_i+e_{66}$, $1\leqslant i\leqslant65$, of the ambient even-weight space, with coordinates $y_i$, and write $e_j(y)$ for their $j$-th elementary symmetric polynomial. The pullback of $\sigma_4$ is
\[
 e_4(y)+e_1(y)e_3(y)=e_4(y)+e_3(y)
\]
as a Boolean polynomial. Thus its quartic part is $e_4(y)$. On this basis the dot product is zero on the diagonal and one off the diagonal. Both sides of~\eqref{eq:polarization} are alternating multilinear forms; they equal $1$ on four distinct basis vectors and vanish when an argument is repeated. This proves the identity after restriction to $C$.

The sum of $h$ in $\mathbb{F}_2$ over an affine four-space equals its fourth derivative along a basis of the direction space. A symplectic basis for a nondegenerate four-space makes the right side of~\eqref{eq:polarization} equal to $1$. If a direction belongs to the radical, the same identity shows that the cubic part of the corresponding derivative vanishes, so that derivative has degree at most two. Affine translation does not change the terms of degree four, which proves the last assertion.
\end{proof}

We use two elementary facts about Boolean polynomials. First, a nonzero polynomial of degree at most $d$ in $n$ variables has at least $2^{n-d}$ ones. This follows by induction from $F=F_0+x_nF_1$. If $F_1\ne0$, the total support of the two slices has size at least the support of $F_1$; if $F_1=0$, the support size doubles. Second, a polynomial of degree less than $n$ has even weight, since each monomial has even weight.

We also need the equality case for quadratic polynomials. For a quadratic Boolean polynomial $Q$ on a vector space, allowing a nonzero constant term, its polar form is $B_Q(u,v)=Q(u+v)+Q(u)+Q(v)+Q(0)$. The same formula applies on an affine space after choosing an origin, and the resulting form on the direction space is independent of that choice. (The earlier form $B$ agrees with this convention because $q(0)=0$.) Splitting off symplectic pairs of the polar form shows the following. A quadratic polynomial of polar rank $2j$ has weight $2^{n-1}$ if its restriction to the radical is nonconstant, and weight
\[
 2^{n-1}\pm2^{n-j-1}
\]
otherwise. Constants and affine-linear polynomials are treated separately. In particular, in seven variables the smallest positive quadratic weights are $32,48,56,64$, and a quadratic of weight $32$ has an affine five-space as its support. In nine variables the minimum positive weight is $128$, attained with support equal to an affine seven-space. These last assertions also follow from the normal form: a quadratic of minimum positive weight is a product of two affine-linear functions with linearly independent linear parts.

\begin{lemma}\label{lem:lowerbounds}
Let $F$ be a Boolean polynomial of degree at most four on an affine space whose direction space carries an alternating form $B$, and suppose that the fourth polarization of $F$ is the right side of~\eqref{eq:polarization}. If the ambient dimension is nine, then
\[
 \begin{aligned}
 \operatorname{rank}B\geqslant6&\quad\Longrightarrow\quad \lvert\{F=1\}\rvert\geqslant64,\\
 \operatorname{rank}B=8&\quad\Longrightarrow\quad \lvert\{F=1\}\rvert\geqslant80.
 \end{aligned}
\]
If $\operatorname{rank}B\geqslant4$, then $F\ne0$.
\end{lemma}
\begin{proof}
The last assertion holds because $F$ has odd weight on a nondegenerate affine four-space. Suppose the rank is at least six. Restricting an alternating form to a codimension-one subspace lowers its rank by at most two. Thus, if the support were contained in an affine hyperplane, the complementary hyperplane would contain a nondegenerate affine four-space disjoint from the support, a contradiction. The same hyperplane argument applies within every affine restriction whose direction space has rank at least six. In particular, on every affine nondegenerate six-space, the support of the restricted function has full affine span. Its weight is even, since its Boolean degree is less than six, and full affine span requires at least seven support points. Hence each such six-space contains at least eight support points. Partitioning the nine-space into eight cosets of one nondegenerate six-space proves the first bound.

Next consider a seven-space of rank six, and let $w$ denote the support size. Two nondegenerate six-slices give $w\geqslant16$. Suppose $w<20$, and let $r_0$ span the radical. If $D_{r_0}F\ne0$, then its degree is at most two and its weight is at most $2w<40$. The quadratic weight classification forces its support to be an affine five-space $A$ of size $32$. Since $D_{r_0}F(x+r_0)=D_{r_0}F(x)$, the space $A$ is invariant under translation by $r_0$. Each of its $16$ pairs $\{x,x+r_0\}$ contributes exactly one point to the support of $F$. If there are no support points outside $A$, that support lies in a proper affine hyperplane, which we have ruled out. Otherwise choose a support point outside $A$. Since $A$ has codimension two, an affine hyperplane through that point can be chosen disjoint from $A$. Its restricted form has rank at least four, so the degree-four minimum-weight bound requires at least four support points there. This is impossible because $w<20$ leaves fewer than four support points outside $A$. Thus $D_{r_0}F=0$. Choose an affine six-space complementary to the radical direction. The descended function on the quotient is represented by the restriction of $F$ to that slice, so its Boolean degree is still at most four. Its weight is therefore even; each support point downstairs has two preimages, hence $4\mid w$ and $w=16$.

Finally, consider a nondegenerate eight-space. Each affine hyperplane has rank six and therefore contains either $16$ or at least $20$ support points. If the total support size $w$ is less than $40$, each pair of complementary hyperplanes has sizes $16$ and $w-16$. Then all $255$ nonconstant Fourier coefficients of the support indicator have absolute value $w-32$, and Parseval's identity would give
\[
 w^2+255(w-32)^2=256w.
\]
Equivalently, $4w^2-259w+4080=0$, which is impossible because its discriminant is $1801$, not a square. Therefore each nondegenerate eight-space contains at least $40$ support points. Its two cosets in the nine-space give the second bound.
\end{proof}

Apply Lemma~\ref{lem:lowerbounds} to $h$ on $V$, using Lemma~\ref{lem:polarization} and $\operatorname{rank}B=9-r$. The cases with $r=3$ have rank six, so $W\geqslant64$, contradicting the upper bounds $32$ and $36$ in~\eqref{eq:Wbounds}. The cases with $r=1$ have rank eight, so $W\geqslant80$, contradicting the upper bounds $52$, $8$, and $78$. The case $r=5$, $S=0$ has rank four, so $h\ne0$, contradicting $W\leqslant0$. Only the following case remains:
\begin{equation}\label{eq:lasttype}
 r=5,\qquad q|_R=0,\qquad S=-128.
\end{equation}

\section{The remaining case}\label{sec:last}

Assume~\eqref{eq:lasttype}. We show that the upper bound $W\leqslant64$ is attained and then derive a contradiction from the resulting equality conditions.

Now $a=64$, $b=4$, and $m\leqslant4$. The set $\mathcal{A}$ is the \emph{annihilator} $\operatorname{Ann}(R)$ of $R$ and has dimension four. Equations~\eqref{eq:gaussweight} and~\eqref{eq:population} become
\begin{equation}\label{eq:lastgauss}
 \sum_{\ell\in\mathcal{A}}\tau_\ell=4,\qquad
 \sum_{\ell\in\mathcal{A}}\tau_\ell(-1)^{\ell(x)}=4(-1)^{q(x)},
\end{equation}
\begin{equation}\label{eq:lastequality}
 4W+4A_{34}+16\beta+32A_{48}+36A_{50}+\mathcal{E}
 =64-48\varepsilon\leqslant256.
\end{equation}
For $x\notin R$, exactly eight elements of $\mathcal{A}$ satisfy $\ell(x)=1$. The two character sums in~\eqref{eq:lastgauss} therefore give
\begin{equation}\label{eq:negativecount}
 \lvert\{\ell\in\mathcal{A}:\tau_\ell=-1,\ \ell(x)=1\}\rvert
 =3+(-1)^{q(x)}.
\end{equation}

\begin{lemma}\label{lem:nonperiodic}
The function $h$ is not invariant under every translation by $R$.
\end{lemma}
\begin{proof}
Suppose $h$ is $R$-invariant. Choose a complement $U$ of $R$ on which $B$ is nondegenerate. By Lemma~\ref{lem:polarization}, $\lvert E\cap U\rvert$ is odd, so
$W=32\lvert E\cap U\rvert$. Since $W\leqslant64$ by~\eqref{eq:lastequality}
and $h(0)=0$, we have $E=x+R$ with $x\notin R$.

Equation~\eqref{eq:radicalm} gives $e(R)\geqslant62$, so the average
weight on $E$ is at most $35$ by~\eqref{eq:supportaverage}.
As $q$ is constant on $E$, it equals $1$ there; otherwise all
weights would be at least $36$. Thus the weights in $E$ are
$34$, $50$, $58$, or $66$, and each word of $E$ contributes at least eight to the left side of
\eqref{eq:lastequality}. Since $8W=256$, equality forces
$\varepsilon=-4$, $\beta=A_{48}=A_{50}=\mathcal{E}=0$,
and every word in $E$ to have weight $34$.
Hence $m=4$, $e(R)=62$, and all $\ell\in Z$ have $\tau_\ell=-1$.
The average weight on $E$ equals $31+z$, where
$z=\lvert\{\ell\in Z:\ell(x)=1\}\rvert$. Thus $z=3$. Since $\tau_\ell=-1$ for every $\ell\in Z$,
\eqref{eq:negativecount} gives $z\leqslant2$, a contradiction.
\end{proof}

We record a consequence of Lemma~\ref{lem:polarization}. For $\ell\in V^*$ with $\ell|_R\ne0$, put $H=\ker\ell$. Since $H+R=V$ and $\operatorname{rad}(B|_H)=H\cap R$, the restricted form has rank four. Each level $\ell=0,1$ therefore contains a nondegenerate affine four-space, so both levels meet $E$ by Lemma~\ref{lem:polarization}.

By Lemma~\ref{lem:nonperiodic}, we may choose $r_0\in R$ with $D_{r_0}h\ne0$. This derivative is quadratic by Lemma~\ref{lem:polarization}, so its support has at least $128$ points. On the other hand, its support is the symmetric difference $E\triangle(E+r_0)$, of size at most $2W\leqslant128$ by~\eqref{eq:lastequality}. Thus $W=64$ and $E\cap(E+r_0)=\varnothing$. The equality case for quadratics then shows that $U_0:=E\cup(E+r_0)$ is an affine seven-space.

Write $L$ for the direction of $U_0$. Then $R\subseteq L$.
Otherwise some $\alpha\in V^*$ vanishes on $L$ but not on $R$.
Since $E\subseteq U_0$ lies in one $\alpha$-level, this contradicts
the preceding observation.

As $R\subseteq L$, every $D_rh$ with $r\in R$ is quadratic and supported in $U_0$. The minimum positive quadratic weight in nine variables is $128=\lvert U_0\rvert$, so $D_rh$ is either zero or $\mathbf{1}_{U_0}$. These alternatives define a linear map $\lambda:R\to\mathbb{F}_2$, because $D_{r+s}h(x)=D_rh(x+s)+D_sh(x)$ and $U_0$ is $R$-invariant. As $\lambda(r_0)=1$, the kernel $P:=\ker\lambda$ has dimension four. The rule $h(x+r)=h(x)+\lambda(r)$ for $x\in U_0$ shows that $E$ meets each of the four $R$-cosets in $U_0$ in exactly one $P$-coset. Equivalently, in adapted affine coordinates with $U_0=\{u_1=u_2=1\}$ and $(z,y)\in R\oplus\mathbb{F}_2^2$,
\[
 h=u_1u_2\bigl(\lambda(z)+Q_0(y)\bigr),\qquad \deg Q_0\leqslant2.
\]

Since $W=64$, equality in~\eqref{eq:lastequality} forces
\[
 \varepsilon=-4,\qquad A_{34}=\beta=A_{48}=A_{50}=\mathcal{E}=0.
\]
Since $A_{42}=0$ and $\mathcal{E}=0$, every word in $E$ has $q=0$ and weight $36$ or $44$. Every nonzero word in $P$ has weight at least $32$, so~\eqref{eq:supportaverage} gives $8e(P)\geqslant15\cdot32$, that is, $e(P)\geqslant60$. The average weight on each of the four $P$-cosets that make up $E$ is at most $66-e(P)/2\leqslant36$. Consequently equality holds throughout: $e(P)=60$, every word in $E$ has weight $36$, and each of the six coordinate functions vanishing on $P$ takes the value $1$ on each of these cosets.

Since $\varepsilon=-4$ and $m\leqslant4$, the set $Z$ consists of four distinct nonzero coordinate functions annihilating $R$. As $P\subseteq R$, all four annihilate $P$ and hence take the value $1$ on $E$. Each $R$-coset in $U_0$ meets $E$, and every $\ell\in Z$ is constant on each such coset. Thus every $\ell\in Z$ is identically $1$ on $U_0=x_0+L$. Such functionals annihilate $L$ and take the value $1$ at $x_0$. Since $\dim\operatorname{Ann}(L)=2$, there are at most two of them, contradicting $\lvert Z\rvert=4$. This excludes~\eqref{eq:lasttype} and completes the proof of Theorem~\ref{thm:main}.

\section{Consequences for dimension ten}\label{sec:dimension10}

\begin{proposition}\label{prop:dimension10}
We have $d_2(122,10)=56$. In addition, the minimum lengths of binary linear codes of dimension ten satisfy
\[
\begin{aligned}
 n_2(10,57)&\geqslant123,& n_2(10,58)&\geqslant124,\\
 n_2(10,59)&\geqslant126,& n_2(10,60)&\geqslant127.
\end{aligned}
\]
\end{proposition}
\begin{proof}
For a binary linear $[n,k,d]$ code with $k\geqslant2$ and $d\geqslant1$, choose a word $u$ of weight $d$ and project onto the $n-d$ coordinates outside its support. The projection has kernel $\langle u\rangle$: otherwise two distinct nonzero codewords $v$ and $v+u$, both supported on the $d$ positions of $u$, would have total weight $d<2d$. Hence the image has dimension $k-1$. If $b$ is the residual weight of a word indexed by $v\notin\langle u\rangle$, then
\[
 \operatorname{wt}(v)+\operatorname{wt}(v+u)=d+2b\geqslant2d,
\]
so $b\geqslant\lceil d/2\rceil$. Therefore
\[
 n_2(k,d)\geqslant d+n_2\bigl(k-1,\lceil d/2\rceil\bigr).
\]
The four minimum-length bounds follow from $n_2(9,29)=66$ and $n_2(9,30)=67$. In particular, the first gives $d_2(122,10)\leqslant56$. Grassl's table records a \emph{quasi-cyclic} binary linear $[121,10,56]$ code, invariant under simultaneous cyclic shifts within its eleven length-$11$ blocks~\cite{GrasslTables}; appending a zero coordinate gives a $[122,10,56]$ code. This establishes the reverse inequality and the stated equality.
\end{proof}

Grassl's table lists the bounds $56$ and $57$ for $d_2(122,10)$~\cite{GrasslTables}; the proposition determines the exact value. For reproducibility, the $[121,10,56]$ code used above has eleven length-$11$ blocks specified by the following generating polynomials in octal notation:
\[
 (2001,365,1001,157,547,327,257,401,555,775,731)_8.
\]
Read each octal integer in binary, with bit $i$ as the coefficient of $X^i$, and work modulo $X^{11}-1$. Take the unshifted blocks and their first nine simultaneous cyclic shifts as generator rows. They are linearly independent, and enumeration of their $2^{10}$ combinations gives the weight enumerator $1+528z^{56}+407z^{64}+88z^{72}$.

The table entries for $[125,10]_2$ and $[126,10]_2$ record maximum-distance upper bounds $58$ and $59$, respectively~\cite{Grassl125,Grassl126}. Thus the last two length bounds in Proposition~\ref{prop:dimension10} recover previously recorded exclusions; they are not asserted to be new code-table improvements.

\providecommand{\bysame}{\leavevmode\hbox to3em{\hrulefill}\thinspace}
\providecommand{\MR}{\relax\ifhmode\unskip\space\fi MR }
\providecommand{\MRhref}[2]{%
  \href{http://www.ams.org/mathscinet-getitem?mr=#1}{#2}
}
\providecommand{\href}[2]{#2}

\end{document}